\documentclass[10pt,journal,compsoc]{IEEEtran}

\usepackage{booktabs} 
\usepackage{setspace, amsmath, amssymb, url, lscape, subfig, algorithmicx, multirow, pslatex, listings, verbatim, alltt, amsfonts, wrapfig, boxedminipage, color, hyperref, bookmark, pifont, algpseudocode}
\usepackage{amsmath}
\usepackage{caption} 
\usepackage{epsfig}
\usepackage{xcolor}
\usepackage{balance}
\usepackage{dblfloatfix} 
\usepackage{enumitem} 

\newcommand{\eg}{{\it e.g., }}
\newcommand{\etal}{{\it et~al., }}
\newcommand{\ie}{{\it i.e., }}

\newcommand{\reviewhl}[1]{#1}

\newlength{\boxfigwidth}

\newcommand{\boxfig}[1]{
\begin{figure}[h]
\begin{center}
\begin{small}
\setlength{\boxfigwidth}{3.1in}
\addtolength{\boxfigwidth}{0in}
\noindent\framebox{\quad\begin{minipage}{\boxfigwidth}
#1
\vspace{-30pt}
\end{minipage} \quad}
\end{small}
\vspace{10pt}
\end{center}
\end{figure}
}

\begin{document}
%
\title{Harnessing the Potential of Function-Reuse in Multimedia Cloud Systems}
\author{%
Chavit Denninnart,~\IEEEmembership{Member,~IEEE,}
Mohsen Amini Salehi,~\IEEEmembership{Member,~IEEE,}

}


\IEEEtitleabstractindextext{%
\begin{abstract}

Cloud-based 
computing systems can get oversubscribed due to the budget constraints of their users or limitations in certain resource types. 
The oversubscription can, in turn, degrade the users perceived Quality of Service (QoS). The approach we investigate to mitigate both the oversubscription and the incurred cost is based on smart reusing of the computation needed to process the service requests (\ie tasks). 
We propose a reusing paradigm for the tasks that are waiting for execution. This paradigm can be particularly impactful in serverless platforms where multiple users can request similar services simultaneously. Our motivation is a multimedia streaming engine that processes the media segments in an on-demand manner. 
We propose a mechanism to identify various types of ``mergeable'' tasks and aggregate them to improve the QoS and mitigate the incurred cost. 
We develop novel approaches to determine when and how to perform task aggregation such that the QoS of other tasks is not affected.
Evaluation results show that the proposed mechanism can improve the QoS by significantly reducing 
the percentage of tasks missing their deadlines 
and 
reduce the overall time (and subsequently the incurred cost) of utilizing cloud services by more than 9\%.

\end{abstract}

\begin{IEEEkeywords}
Task Aggregation, Over-subscription, Serverless Computing, Cloud computing, Video Stream Processing.
\end{IEEEkeywords}

}
\markboth{}%
{}

\maketitle
\IEEEdisplaynontitleabstractindextext
\IEEEpeerreviewmaketitle

\IEEEraisesectionheading{\section{Introduction}
\label{sec:intro}
}


Serverless computing or Function-as-a-Service (FaaS) is gaining popularity as an on-demand and cost-efficient computing solution for cloud-based applications. \reviewhl{A serverless computing system is defined as a cloud-based computing system that can execute functions (tasks) without involving the user in the server management. This high-level transparency provides  the illusion that there is no server in place.} Modern software engineering practices, such as DevOps \cite{ebert2016devops} and Continuous Integration Continuous Delivery (CI/CD) \cite{meyer2014continuous}, operate based on splitting an application into several micro-services \cite{lloyd2018serverless} where each microservice can be hosted by the serverless computing platform. 

Behind the scene, the serverless computing platform seamlessly handles the resource allocation and execution of the micro-services on the cloud resources. A common practice is to \reviewhl{serve multiple users' micro-services (a.k.a. \emph{task requests}) on the provider's shared scheduling queue.} The tasks often have individual deadlines that failing to meet them compromises the Quality of Service (QoS) expected by the end-users. The platform's scheduler allocates the tasks to an elastic pool of computing resources such that their QoS expectations are fulfilled. 


\emph{Oversubscription} in a computing system is defined as a system that is overwhelmed with arriving tasks such that it is impossible to satisfy the users' QoS expectations. In particular, the serverless computing platforms are prone to oversubscription for the following reasons:
(A) Even though clouds supply virtually unlimited resources, users generally have budget constraints, and they cannot lavishly acquire cloud resources \cite{bi2017application}; 
(B) Privately hosted serverless computing platforms (and those used in fog/edge systems) fall short on the elasticity and scalability aspects \cite{ling2019pigeon}; 
(C) Depending on the function trigger events, the tasks arrival pattern is often uncertain and includes surges \cite{facebookvideolive18}; and 
(D) To maximize their profit, cloud providers tend to increase the number of tasks served on the minimum number of machines.

A large body of research has been dedicated to mitigating the oversubscription problem 
in the computing system. The approaches undertaken in these research works follow two main lines of thinking: \emph{First}, resource allocation based approaches (\eg \cite{liu2016dynamic,alfayly2015qoe,hou2015qoe}) that try to minimize the impact of oversubscription through efficient mapping (scheduling) of the task tasks to the resources. \emph{Second}, approaches based on the computational reuse (\eg \cite{zhang2012distributed,casas2017balanced}) that avoid or mitigate the oversubscription through efficient caching of the computational results. The latter is particularly effective when there is a redundancy in arriving tasks. 

Although both of the aforementioned approaches are effective, they are limited in certain ways. The allocation-based approaches mitigate the impact of oversubscription but cannot entirely resolve it, according to the definition of the oversubscription. In addition, many of the approaches are based on complex scheduling algorithms that impose extra overhead to the already overwhelmed system~\cite{guo2015workflow}. The reusing approaches that operate based on caching are also limited because they can only reuse the computations for tasks that are identical to the ones already completed and cached~\cite{zhang2014efficient}. In other words, if two tasks share part of their computation, caching cannot reuse the result of the shared part~\cite{andrade2019optimizing}.

\begin{figure}[b]
\vspace{-10pt}
\centering
\includegraphics[width=0.45\textwidth ]{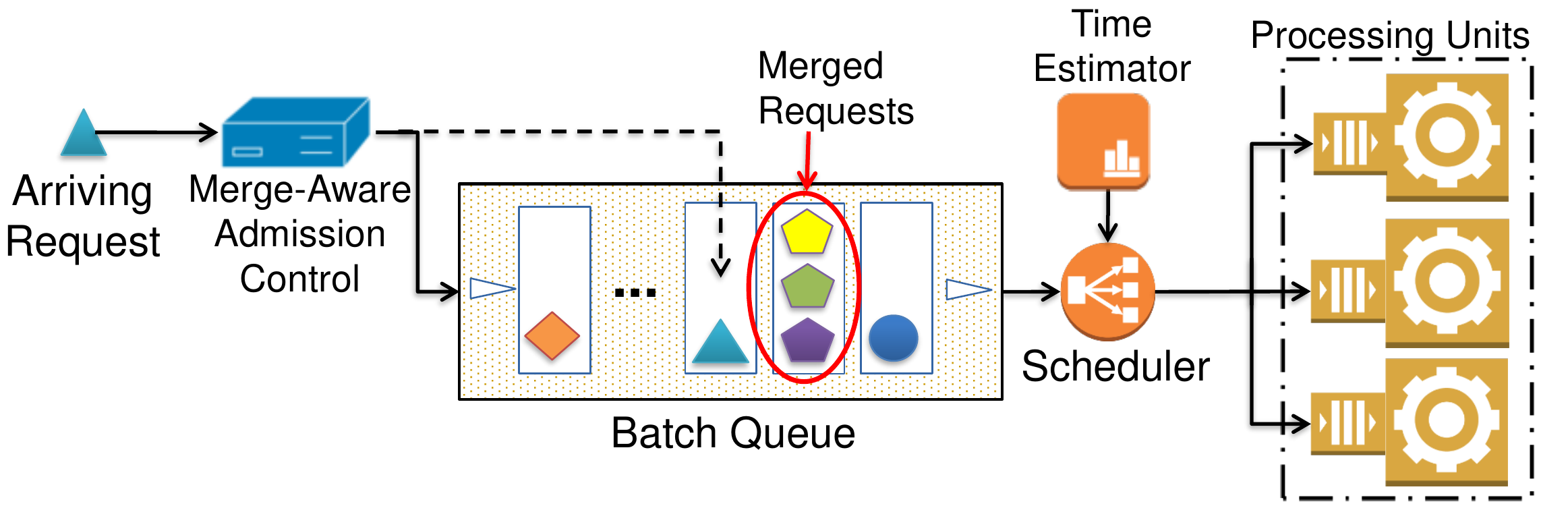}
    \caption{\small{The overview of the task aggregation procedure. \reviewhl{A new task arrives at the Admission Control can be merged with an existing one in the Batch Queue}. Task shapes represent different task types in the system, and shape color represents different task configurations.}}
\label{fig:arch} 
\end{figure}

Accordingly, in this research, we propose a mechanism based on the computational reuse approach that aims at alleviating oversubscription by aggregating similar tasks in the task scheduling queue of the \reviewhl{cloud platforms. Such scheduling queue for shared resource is particularly common behind the scenes of serverless computing platforms. However, it is also applicable to other systems.} As shown in Figure~\ref{fig:arch}, the mechanism can aggregate (\ie merge) not only identical tasks but also those that partially share their computation. We note that our mechanism complements existing allocation- and caching-based approaches and is not a replacement for them. In fact, the merging mechanism makes the scheduling queue less busy and potentially lighten up the scheduling process. Caching-based approaches are also complemented by capturing the in-progress tasks and those that are partially similar.

To reuse part of the computation, a question that needs to be addressed is how to identify \emph{mergeable} tasks? An arriving task can potentially have multiple mergeable pairs with varying levels of similarity. Also, the solution for task similarity detection should not impose an extensive overhead on the system. The other concern in merging tasks is to form large compound tasks that potentially cause missing the deadline of either the merged tasks or other pending tasks waiting behind the merged task. As such, merging tasks raises the following two problems: (A) \emph{What are different types of mergeable tasks and how to detect them?} (B) \emph{How to perform merging without endangering other tasks in the system? 
}


Our motivation in this research is a multimedia streaming engine that needs to process multimedia contents (\eg changing resolution or bit-rate of video) in a serverless cloud platform before streaming them to viewers~\cite{CVSSJournal}. Multiple viewers can stream multimedia contents in various configurations, hence, creating similar or identical tasks in the system. In particular, when the system is oversubscribed, the likelihood of having mergeable tasks increases. 
In this context, our proposed mechanism can detect identical and similar tasks and reuse the whole or part of the computation by merging them. Intelligently achieving task merging can benefit both the viewers, by enabling more tasks to meet their deadlines, and the stream providers by improving resource utilization and reducing their incurred cost of using services. 



In this research, we develop an Admission Control module (see Figure \ref{fig:arch})
that detects different levels of similarity between tasks and performs merging by considering the tasks' deadlines. 
In summary, the \textbf{key contributions} of this research are as follows:
\begin{itemize}
 \item Proposing an efficient method to identify mergeable tasks.
 \item Proposing methods for proper positioning of merged tasks in the scheduling queue.
 \item Determining appropriateness and potential side-effects of merging tasks considering the oversubscription level of the system.
 \item Analyzing the performance of merging on the viewer's QoS and the cost of utilizing the processing units.
\end{itemize}

Although we develop this mechanism in the context of the multimedia streaming system on a serverless platform, the idea of task aggregation and research findings of this work are valid for other domains. However,
we note that identifying mergeable tasks is domain-specific and requires task profiling for each particular system.

The rest of the paper is organized as follows: 
Section \ref{sec:background} provides some background on multimedia stream processing engine; 
In Section \ref{sec:arch} we present an overview of the proposed design; 
In Section \ref{sec:detection} we propose an efficient method to identify mergeable tasks detection; 
In Section \ref{sec:appropriateness}, we present merge appropriateness determination and merge position finder; 
In Section \ref{sec:adapt}, we discuss ways to quantify oversubscription levels. As well as ways to use quantified oversubscription level to enhance tasks merging decision.
In Section \ref{sec:evltn}, we perform performance evaluations; In Section \ref{sec:relwk} we discuss related works; And finally we conclude the paper and provide potential future works in Section \ref{sec:conclsn}.

\section{Background}
\label{sec:background}

\subsection{Serverless Computing Systems}
In the serverless computing system, a serverless application is composed of one or more stateless standalone microservices that handle specific types of service request~\cite{peekserverless}. 
In this computing model, the resource provider manages all execution environments such as resource allocating, scaling, scheduling, and ensuring availability. To use serverless computing, the user uploads the micro-services (functions) that handle service requests that can be triggered by a timer or an event (\eg HTTP request). In the back end, frequently used functions are maintained in the memory to enable a fast (warm) start of the function upon arrival of a service request~\cite{shahrad2020serverless}. As this treatment is not affordable for infrequently used functions, infrequently used function is removed from the memory and have to launch from a cold state upon invocation. 
Each service request is submitted to a shared scheduling queue for execution. As such, in comparison with the conventional IaaS cloud, serverless platforms reduce resource idling; hence, improve the cost-efficiency. 




\subsection{Multimedia Stream Processing}

Multimedia (\eg video) contents, either in the form of on-demand streaming or live streaming, usually have to be processed before streaming them to the viewers. A wide range of processing---from object detection \cite{redmon2018yolov3} to changing compression standards \cite{artigas2007discover}---can be applied to the multimedia segments. A common type of processing on video segments that we consider in this study is to convert, a.k.a transcode, them to match the characteristics of clients' display devices~\cite{CVSS,amini20videosurvey}. Transcoding can encompass operations such as \emph{bit rate adjustment, spatial resolution reduction, frame rate reduction}, and \emph{compression standard (codec) conversion}. Particularly, for live streaming, video segments have to be transcoded upon arrival~\cite{vlsc}. 

A multimedia stream 
 is composed of several media segments of varying or fixed length. While varying the segment length based on the content can be more space-efficient, in practice, \reviewhl{most existing video streaming systems use a fixed-length segment of around 2 seconds for simplicity}. 
 Each segment processing request is considered as a task that has an individual deadline \cite{CVSSJournal,matin_paper}. Deadline violation of any task reduces the Quality of Service (QoS) perceived by the viewer. 

\subsection{Serverless Multimedia Streaming Engine (SMSE)}
We develop our mechanism within the context of Serverless Multimedia Streaming Engine (SMSE) that enables on-demand (\ie lazy) processing of multimedia streams~\cite{CVSSJournal}. Even though SMSE can be fed by any type of user-defined multimedia processing function, in this study, we use it for the common case of video transcoding functions. 

\begin{figure}[ht]
\centering
    \includegraphics[width=0.36\textwidth]{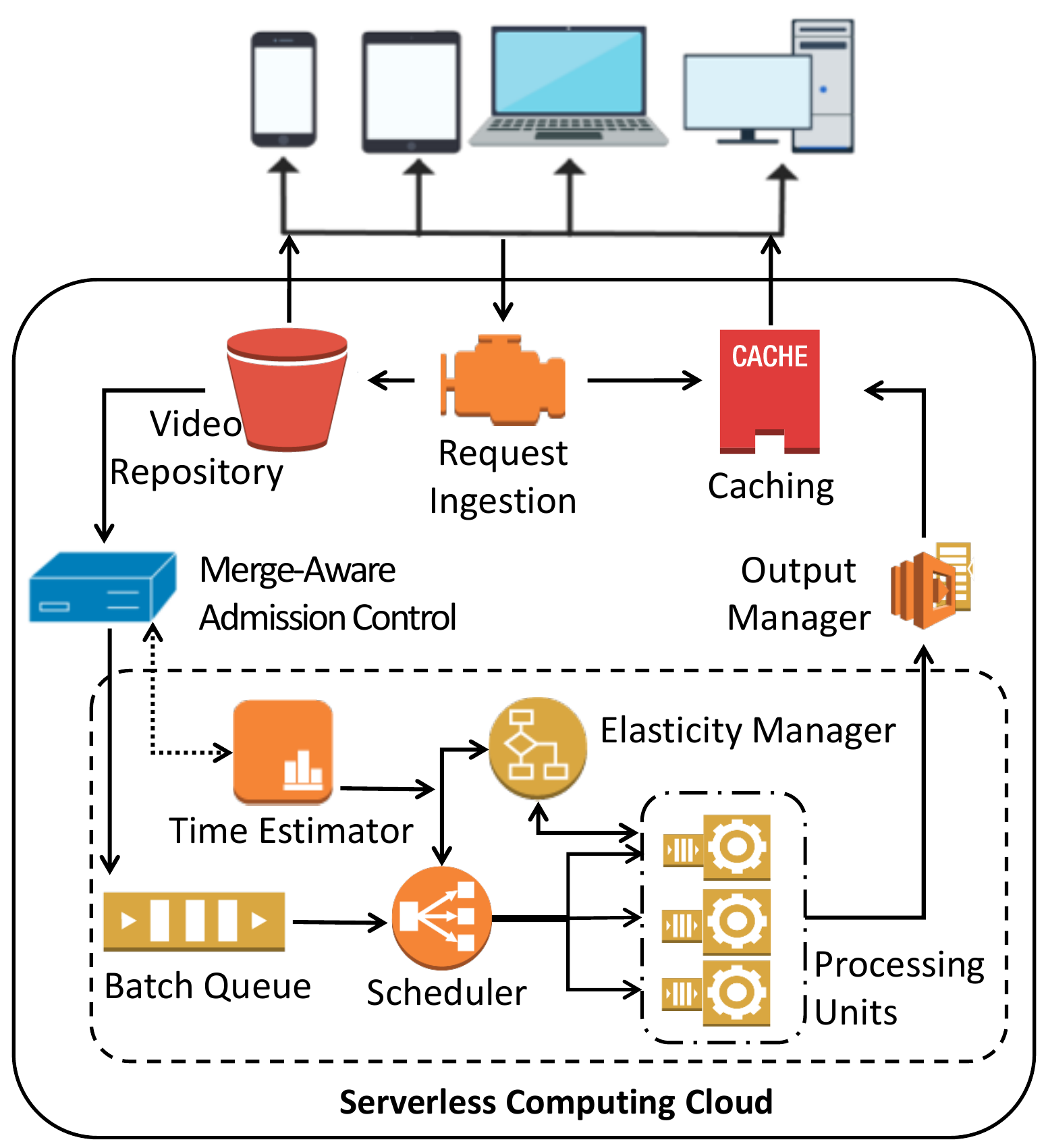}
    \caption{\small{Bird-eye view of the serverless multimedia streaming engine (SMSE) platform that is used in our study to the process of multimedia streams.}}
    \label{fig:CVSE}
    \vspace{-10pt}
  \end{figure}
  
In the SMSE architecture, as shown in Figure~\ref{fig:CVSE}, the Splitter dispatches segments of the requested multimedia stream. A segment processing request includes the operation required along with the corresponding parameters bound to that task. Each arriving task request is assigned an individual deadline and a priority by the Admission Control component. Then, the Admission Control sends the task to the shared batch queue where the request waits to be assigned by the Scheduler to one of multiple processing unit's queues. Scheduler's batch queue is managed based on a certain scheduling policy, such as FCFS, Earliest-Deadline-First (EDF), Max-Urgency-First, and Highest-Priority-First~\cite{matin_paper}. Most of the scheduling policies rely on the Time Estimator component to provide the expected execution time of each task type (\ie different transcoding functions) on a given machine type. Such estimation can be obtained based on historical execution time information of different transcoding functions~\cite{CVSS}.

Once a task is in the execution queue, its function and required data (that is, the multimedia segment itself) are fetched for execution. Output Manager orders the processed segments and transmits them to the viewer. Segments of the multimedia that are getting popular are recognized by the Caching component of SMSE and are stored to enable caching-based computational reuse.

SMSE often receives different forms of \emph{mergeable} task. For example, two viewers who use similar display devices may request to stream the same content. 
Alternatively, two viewers with dissimilar display devices (\eg different resolution and compression standard) or personal requirements (\eg audio translation or graphic censorship) may stream the same video, but with different specifications. The former case creates \emph{identical} tasks in the system, whereas the latter one creates \emph{similar} tasks. 

We develop our task merging mechanism inside the Admission Control component of SMSE. Upon task arrival, Admission Control recognizes if it is mergeable with the existing tasks in the batch queue.
Then, it decides if the arriving task can and should be merged with the existing task or not. 
Before these steps, the Caching component also verifies if the transcoded format of the requested segment is already cached.

\section{Overview of the Admission Control Mechanism to Reuse Computation via Task Merging}
\label{sec:arch}


Admission Control is the front gate of the batch queue, and it is in charge of performing merging arriving tasks with the ones already in the batch queue. The reason we do not perform the merging in the batch queue (\ie after the task admission) is that, in that case, to find mergeable tasks, we need to scan the entire queue and perform a pair-wise matching between the queued tasks, which is inefficient and implies a significant number of redundant comparisons.  

\begin{figure}[t]
\centering
    \includegraphics[width=0.45\textwidth]{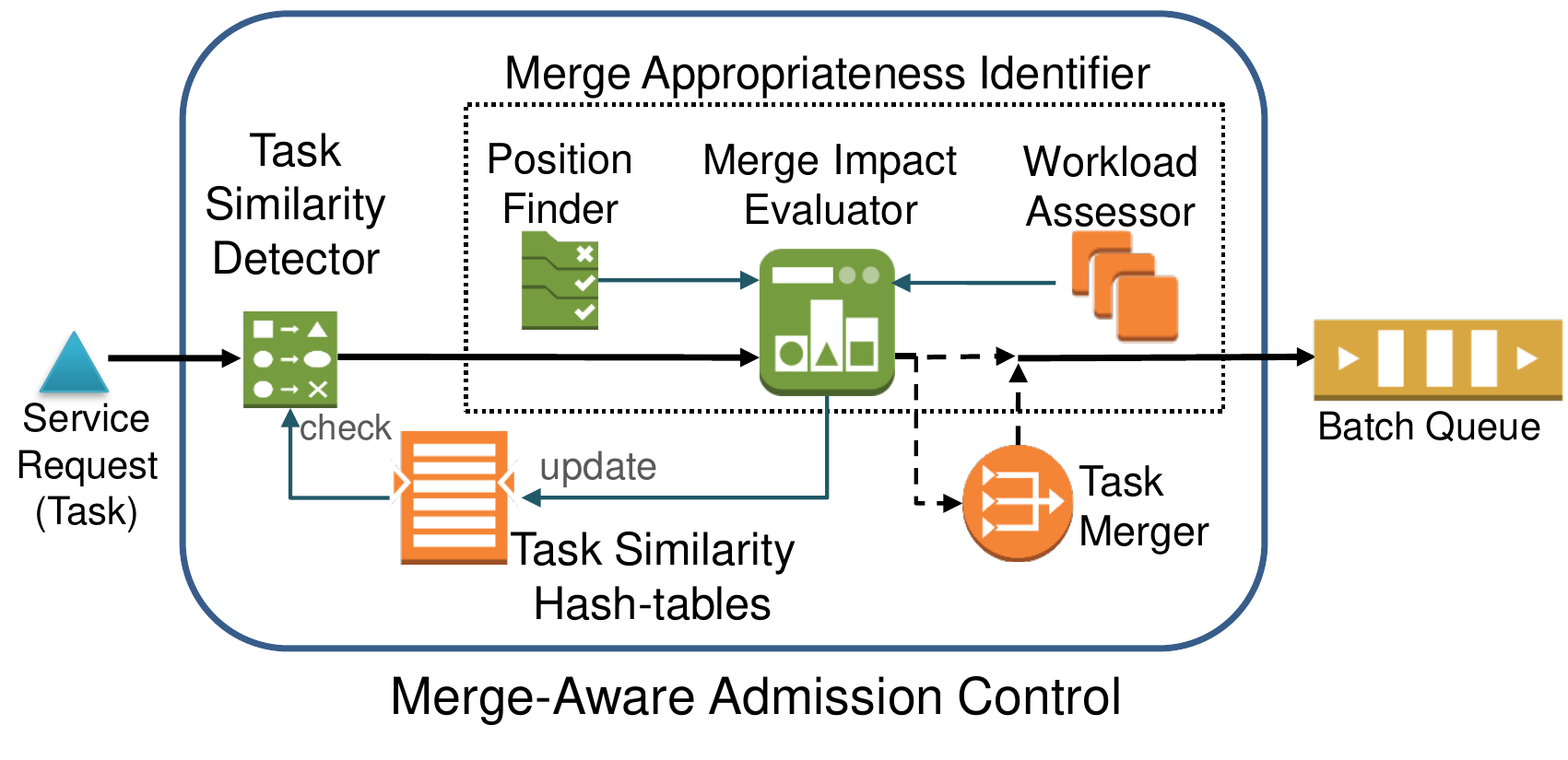}
    \caption{\small{Task aggregation mechanism inside Admission Control of SVSE. Before adding a task to the batch queue, it is checked if it is mergeable with any other queued tasks and whether or not the merging operation is appropriate to be achieved.} }
    \label{fig:adc}
    \vspace{-10pt}
  \end{figure}
  
The proposed task merging mechanism, shown in Figure \ref{fig:adc}, consists of three main components as follow\reviewhl{s}: (A) Task similarity detector; (B) Merging appropriateness identifier; and (C) Task merger. 
\textbf{Task Similarity Detector} is a lightweight method based on hashing techniques to identify mergeable tasks. As detailed in Section \ref{sec:detection}, it maintains multiple hash tables to cover multiple levels of tasks' mergeability. If the arriving task is identified mergeable with an existing task, then the system employs the \textbf{Merge appropriateness identifier} to assess if performing the merge on the identified tasks can impact other tasks in the system or not. Merge appropriateness identifier has three cooperating modules. \textit{Position Finder} locates the suitable position for merged tasks in the scheduling queue, such that the other tasks are not affected. To examine each position, Position finder consults with \textit{Merge Impact Evaluator} to estimate which and how many tasks can potentially miss their deadlines as the result of merging. The task merging decision is made based on system oversubscription level obtained from \textit{Workload Assessor} (see Section~\ref{sec:appropriateness} and~\ref{sec:adapt} for further details). Once the merging is confirmed as appropriate in a certain position of the batch queue, \textbf{Task merger} component carries out the merge operation on the two tasks. 




\section{Task Similarity Detection}

\label{sec:detection}

\subsection{Categories of Mergeable Tasks}
Mergeability of two given tasks can be explained based on the amount of computation the two tasks share. In particular, mergeability of two or more tasks can be achieved in the following levels: 
\begin{enumerate}[label=(\Alph*)]

 \item \emph{Task level}: This is when more than one task which creates the same processing task present to the scheduling queue. Therefore, this level is also known as 
 \emph{Identical tasks} and can achieve maximum computational reusability. For instance, consider two viewers without  personalized requirements stream the same content and need it to be transcoded with the same resolution to be displayed on compatible devices. As these tasks lead to identical multimedia processing, merging them consumes the same resources required for only one task, hence, reducing both cost and processing delay.


 \item \emph{Data-and-Operation level}:  
This is when two or more tasks perform the same operation on the same data but with different configurations. \reviewhl{The combined processing task has equivalent results as if each task is processed individually.}
Computational reusability can be achieved through the sharing of function loading overhead and common processing steps. For instance, consider two viewers who stream the same video content with different resolutions. Without merging, the two tasks need to load the video, decode it, and encode it separately. However, by merging the two tasks, the loading and decoding operations can be shared, then the encoding operation is carried out separately.  
 

\item \emph{Data-only level}: 
This is when the only common sharing specification between the two (or more) tasks is only on the data. Tasks that share the same data can reduce the data retrieval overhead. 
This third tier task similarity level saves the least amount of processing time in comparison to other cases.

 \end{enumerate}

It is noteworthy that, although merging increases the execution time of the merged task (except in the Task level), our other study \cite{shangrui_paper} shows that the execution time of the merged task is remarkably (up to 40\%) shorter than the combined execution time of the unmerged tasks.
\reviewhl{Also, the aforementioned reusability forms are context-specific and can be re-designed to fit other contexts. This is particularly the case for the third level (data-only) task similarity. In the video processing context, the overhead of loading data (\ie video contents) is higher than the containerized function \cite{davood20}. Therefore, we prioritize merging tasks with the same data, rather than the same function. In other contexts (\eg big data analytics and machine learning), there can be potentially more benefits to group tasks based on the same operation (function), rather than the same data.
}

In SMSE, the Admission Control component can achieve task level reusability for the same segments that need to be processed with the same function, but for different viewers. Data-and-Operation level reusability is achieved for segments that perform the same transcoding function but with different configurations. Finally, Data-only reusability is achieved for the same segments that are served by different functions.

\subsection{Detecting Similar Tasks}\label{sec:mergable-detection}

Assuming there are $n$ tasks in the queue, for each arriving task, a na\"{i}ve mergeable task detection method has the overhead of performing $n$ comparisons to find the mergeable tasks. To reduce the overhead, we propose a method that functions based on hashing techniques. The general idea of the proposed method is to generate a hash key from the arriving task request signature (\eg multimedia segment id, processing type, and their parameters). Then, the Admission Control finds mergeable tasks by searching for a matching key in the hash table of existing tasks in the scheduling queue. 

%
The explained method can detect Task level mergeability. We need to expand it to detect other levels as well. To maximize the computational reusability, an arriving task is first verified against Task level mergeability. If there is no match in the Task level, then the method proceeds with checking the next levels of mergeability, namely Data-and-operation level and Data-only level, respectively. To achieve the multiple levels of mergeability, we create three hash-tables---each covers one level of mergeability. The hash-keys in each level are constructed from the tasks' characteristics that are relevant in deciding mergeability at such level. For instance, in the multimedia streaming case study, keys in the hash-table that verifies task level mergeability are constructed from media segment id, processing type, and their parameters. Alternatively, keys in the hash-table that verifies Data-and-operation level mergeability are constructed from media segment id and processing type. Similarly, keys in the hash-table of Data-only level mergeability are constructed from segment id. 

\boxfig{
Upon arrival of task $j$: 
\begin{itemize}

\item[(1)] if $j$ merges with existing task $i$ on Task level similarity:
\begin{itemize}
\item[--] No update on hash-table is required
\end{itemize}
\item[(2)] if $j$ merges with existing task $i$ on operation-and-data or Data-only level similarity:
\begin{itemize}
\item[--] Add an entry to each hash-table with hash-keys of task $j$ and point them to merged task $i+j$ 
\end{itemize}
\item[(3)] if $j$ matches with existing task $i$ but the system chooses not to merge them:
\begin{itemize}
\item[--] Add an entry to each hash-table with hash-keys of task $j$ and point them to task $j$
\end{itemize}
\item[(4)] if $j$ does not match with any of the existing tasks:
\begin{itemize}
\item[--] Hash-keys of task $j$ are added to the respective hash-tables 
\end{itemize}
\end{itemize}

Upon task $j$ completing execution (\ie dequeuing task $j$):
\begin{itemize}
\item[--] Remove all entries pointing to task $j$ from hash-tables
\end{itemize}
\caption{\small{The procedure to update hash-tables upon arrival or completion of tasks.}} 
\label{fig:htupdate}
\vspace{5pt}
}

Each entry of the hash-tables includes a hash-key and a pointer to the corresponding task. Entries of the three hash-tables must be updated upon a task arrival and execution. The only exception is Task level merging, which does not require updating the hash-tables. Figure \ref{fig:htupdate} shows the procedure for updating the hash-tables for a given task $j$. 

When the system merges task $j$ with existing task $i$, the merged task, denoted as $i+j$, is essentially the object of task $i$ that is augmented with task information (\eg processing parameters) of task $j$. In this case, as shown in Step (2) of this procedure, the system only adds an entry to each hash-table with the hash-key of task $j$ pointing to merged task $i+j$. This is in addition to the entry for task $i$ pointed to task $i+j$ that already exist in the table. 
When task $j$ is mergeable with existing task $i$, but the system decides to add task $j$ to the batch queue without merging. 
In this case, \reviewhl{it suggest that task $i$ has  certain characteristics (\ie tight deadline) which make it unsuitable for task merging. Meanwhile, Task $j$ is freshly arrived and has a higher likelihood of merging with other arriving tasks.} Hence, as shown in Step (3) of the procedure, the matching entry pointing to task $i$ is redirected and points to task $j$. 
It is worth noting that if the arriving task does not match with any of the existing tasks, as shown in Step (4), its hash-keys must be generated and added to the respective hash-tables. Also, when a task is served (processed), its corresponding entries are removed from the hash-tables.


\section{Identifying Merging Appropriateness}
\label{sec:appropriateness}
\subsection{Overview}
Assume that arriving task $j$ has Data-and-operation or Data-only similarity with existing task $i$. Also, assume that task $i$ is scheduled ahead of at least one other task, denoted task $k$, in the scheduling queue. Merging task $j$ with $i$ either delays the execution of task $k$ or task $i$. Such an imposed delay can potentially cause task $k$ or $i$ to miss their deadlines. Therefore, it is critical to assess the impact of merging tasks before performing the merge. 
The merge should be carried out only if it does not cause more QoS violations than it improves.
It is noteworthy that Task level merging does not delay the execution of other tasks; thus, it can always be performed.

Accordingly, in this section, we first introduce the Merge Impact Evaluator component, whose job is to assess the impact of the merging arriving task on existing tasks. Later, we introduce Position Finder, whose job is to position the arriving task in the scheduling queue, either through merging with other tasks or as a new entry in the scheduling queue.

\subsection{Evaluating the Impact of Merging}
\label{sec:impact}
Ideally, task aggregation should be performed without causing deadline violations for other tasks. Accordingly, the impact of merging two or more tasks is evaluated based on the number of tasks missing their deadlines due to the merging. \reviewhl{The evaluation requires the Time Estimator component (see Figure~\ref{fig:arch}) to estimate the execution time of the tasks. In the video processing context, it is proven that the execution time of the tasks either follows Normal distribution or can be approximated by it~\cite{CVSSJournal,vlsc,matin_paper,shangrui_paper,li2018performance}. Accordingly, we employ Normal distribution to model the execution time, which is less compute-intensive than other discrete and continuous distributions.}
To evaluate the impact of merging, a temporary structure, called \textit{virtual queue}, is constructed that contains a copy of machine queues. Then, we assume the merging has taken place on the tasks in the batch queue and schedule them to the virtual queue according to the scheduling policy. This enables us to estimate the number of tasks missing their deadlines in the presence of merging.

To assure the minimal impact of the merging, by default, worst-case analysis is performed on the completion time of the affected tasks to estimate the number of tasks missing their deadlines. For a given task $i$, we assume its execution time follows a Normal distribution~ \cite{CVSSJournal,li2018performance,hussain2019federated} and $\mu_i$ and $\sigma_i$ represent the mean and standard deviation of its execution time. Let $E_i$ be the estimated execution time of task $i$. In the worst-case analysis, we consider $E_i$ to be large enough that with a high probability (97.7\%), the real execution time is less than $E_i$. As such, $E_i$ is formally defined based on Equation~\ref{eq:e_i}.

\begin{equation}\label{eq:e_i}
E_i=\mu_i + \alpha\cdotp \sigma_i    
\end{equation}
 
\reviewhl{
In this equation, $\alpha$ is the standard deviation coefficient, and its default value is 2. That implies with 97.7\% chance of task $i$ will not missing its deadline because of the merging. Note that to encourage more aggressive merging under oversubscription, we can relax the pessimissity of the worst-case analysis by diminishing the value of  $\alpha$ (see Section \ref{sec:adapt} for further details).
}

Once we know $E_i$, we can leverage it to estimate the completion of task $i$ on a given machine $m$, denoted as $C_i^m$. We know that calculating $C_i^m$  involves the summation of the following four factors: (A) current time, denoted $\tau$; (B) estimated remaining time to complete the task currently executing on machine $m$, denoted $e_r^m$; (C) sum of the estimated execution times of $N$ tasks that are pending in machine queue $m$, ahead of task $i$. This is calculated as $\sum_{p=1}^{N}(\mu_p + \alpha\cdotp \sigma_p)$; (D) estimated execution time of task $i$. 
The formal definition of $C_i^m$ is shown in Equation~\ref{eq:compl},

 \begin{equation}\label{eq:compl}
  C_i^m = \tau + e_r^m + \sum_{p=1}^{N} (\mu_p +\alpha\cdotp \sigma_p)  + (\mu_i + \alpha\cdotp \sigma_i)
 \end{equation}

In the tie situation that the number of tasks missing their deadlines with and without merging is the same, we choose to perform merging to reduce the overall time of using cloud resources. However, one may argue an alternative approach to not perform task merging because merging can marginally increase the chance of missing deadlines for other tasks. 

\subsection{Positioning Aggregated Tasks in the Scheduling Queue}

Once two tasks are detected as mergeable, the next question is: where should the merged task be placed in the batch queue? The number of possible answers depends on the scheduling policy of the underlying serverless computing platform. 
If manipulating the order of tasks in the batch queue is allowed, then the Position Finder examines possible locations for the merged tasks in the queue. For each location, it consults with the Merge Impact Evaluator component (see Figure \ref{fig:adc}) to identify if the merge has potential side-effects on the involved tasks or not. Once Position Finder locates an appropriate position, it notifies Task Merger to construct the merged task.

Scheduling policies usually sort tasks in the batch queue based on a certain metric (known as the queuing policy). For instance, Earliest Deadline First~\cite{CVSS} sorts the queued tasks based on their deadlines. This assumption restricts the number of positions can be identified for the merged tasks that in turn limits the performance gain of task merging. 
To conduct a comprehensive study, in this section, we investigate two main scenarios: (A) when the queuing policy is mandated (elaborated in Sub-section~\ref{sec:posrespect}); (B) when the queuing policy is relaxed (elaborated in Sub-section~\ref{sec:posignore}). 



\subsubsection{Task Positioning while Queuing Policy is Maintained}
\label{sec:posrespect}
In this part, we study three commonly used queuing policies: \textbf{(a)} First Come First Served (FCFS); \textbf{(b)} Earliest Deadline First (EDF); and \textbf{(c)} Max Urgency. While FCFS and EDF are known queuing policies, Max Urgency sorts the tasks in the queue based on tasks' deadline and execution time. More specifically, for task $i$, urgency is calculated as $U_i=1/(\delta_i - E_i)$, where $U_i$ is urgency score of task $i$, $\delta_i$ is its deadline, and $E_i$ is its estimated execution time. 

\paragraph{\textbf{FCFS}} Let $j$ be the arriving task and $i$ a matching task already exists in the queue. We can merge tasks by either augment task $i$ with $j$'s specification or cancel task $i$ and reinsert $i+j$ into the queue. Therefore, the arrival time of the merged task ($i+j$) can be either the arrival time of task $i$ or task $j$. In the former case, $i+j$ delays the completion time of all tasks located behind $i$. In the latter case, $i+j$ only delays completion time of $i$. In either case, the delayed task(s) can potentially miss their deadline(s) due to the merge operation. A compromise between these two extreme positions is possible and is described in Sub-section \ref{sec:posignore}.

\paragraph{\textbf{EDF}} In this policy, tasks with an earlier deadline are positioned earlier in the queue. When two or more tasks are merged, each of them still keeps its individual deadline. However, only the earliest deadline is considered for the queuing policy. Assuming that task $i$ has an earlier deadline than $j$, task $i+j$ can be only positioned in task $i$'s spot.

\paragraph{\textbf{Max Urgency}}
Recall that except in Task level merging, other levels of merging increase the execution time of the merged task. In this case, the urgency of $i+j$ is: $U_{i+j}=1/(min(\delta_{i},\delta_{j})-E_{i+j})$. This means the urgency of the merged task is increased. Thus, the merged task can potentially move forward in the queue and get executed earlier. As such, tasks merging in max urgency queue can potentially cause missing the deadline of tasks located ahead of $i$ in the scheduling queue as well.

\subsubsection{Task Positioning while Queuing Policy is Relaxed}
\label{sec:posignore}
Queuing policies mentioned in the previous part are not aware of task merging. Except for Max Urgency that moves the merged task forward in the queue due to the increase in the merged task urgency, other policies do not relocate the merged task. However, a more suitable position for the merged task can be found by relaxing the queuing policy. In this case, assuming there are $n$ tasks in the batch queue, the merged task, $i+j$, has to be examined against $n+1$ possible locations to find the position that maximizes the chance of all tasks meeting their deadlines. Examining each possible location implies evaluating the impact of merging, hence, calling the scheduling method. Assuming there are $m$ machines in the system, each impacts evaluation costs $n\cdotp m$ and performing such evaluation for all $n+1$ possible locations implies $(n^2+n)\cdotp m$ complexity. \reviewhl{This makes the time complexity of finding an optimal solution as approximately $O(m \cdotp n^2)$}.

Such overhead itself is a burden to the system that is already oversubscribed. As such, in the rest of this section, we propose two Position Finding heuristics and analyze them. The objectives of these heuristics are: first, not to allow the merged task to miss its deadline; and second, do not cause other tasks to miss their deadlines.

\paragraph{\textbf{Logarithmic Probing Heuristic}} This heuristic evaluates the impact of merging when $i+j$ is in the middle of the queue. The evaluation result dictates how to proceed with the probe as follows:

(i) The position neither causes deadline violation for other tasks nor  $i+j$. Therefore, the appropriate position is found.

(ii) Task $i+j$ misses its deadline, but the number of other tasks missing their deadlines does not increase as a result of merging. This implies that $i+j$ should be executed earlier. Thus, the procedure continues to probe in the first half of the queue. 

(iii) Task $i+j$ meets its deadline, but the number of other tasks missing their deadlines increases as a result of merging. This implies that $i+j$ should be executed later to reduce the merging impact on other tasks. Thus, the procedure continues to probe in the latter half of the queue.

(iv) Task $i+j$ misses its deadline, and the number of other tasks missing their deadlines increases as a result of merging. Then, stop the procedure and cancel merging because the procedure cannot find an appropriate position for merging.

The aforementioned steps are repeated until it terminates, or there is no position left to be examined in the batch queue. In the latter case, we stop the procedure and cancel merging.

\paragraph{\textbf{Linear Probing Heuristic}} 
In the FCFS policy, we know that the order of tasks in the batch queue implies the order of their execution. That is, placing a task in position $p$ of the queue only delays tasks located behind $p$. That said, the \emph{first} phase of this heuristic aims at finding the latest position for task $i+j$ in the batch queue so that it does not miss its deadline. The latest position for $i+j$ in the queue implies the minimum number of tasks are affected ---those located behind the merged task. 

To carry out the first phase, the procedure constructs virtual queues to find the latest position for $i+j$. For that purpose, it alternates the position of $i+j$ in the batch queue, starting from the head of the queue. In each position, the completion time of $i+j$ is calculated based on the tasks located ahead of it and is examined if $i+j$ misses its deadline. Once task $i+j$ misses its deadline, the previous position is the latest possible location for it not to miss its deadline.

Once we found the latest position for $i+j$, we need to verify if the insertion of $i+j$ causes any deadline violation for the tasks behind it or not.
For that purpose, in the \emph{second} phase, we only need to evaluate the merging (via Merging Impact Evaluator) once. If there is no impact, then the found position is confirmed. Otherwise, the merging is canceled.
\reviewhl{
 It is noteworthy that this procedure is efficient because the virtual queue is created only once. Also, after each task assigned to the virtual queue, it simply adds one more condition to check $i+j$ completion time. }

\paragraph{\textbf{Analysis of the Heuristics}}

In this part, we analyze Logarithmic Probing and Linear Probing heuristics in terms of their complexity and optimality of the position they find. 

\emph{Complexity Analysis.} 
Phase one of Linear Probing Heuristic examines $n$ tasks to be scheduled on $m$ machines with an additional check if $i+j$ can be scheduled on time directly after each of the $n$ tasks. That results in $n\cdotp m$ complexity to provide a single position for Phase two to verify. Phase two is essentially evaluating the impact of merging, which again needs $n$ tasks to be scheduled on $m$ machines. The combined complexity of the two phases is $2\cdotp n\cdotp m$. Alternatively, Logarithmic Probing Heuristic spends trivial computation of $O(1)$ to pick a position in the batch queue to verify the appropriateness. If the position is identified as inappropriate, the search continues for up to $\log n$ positions. Since the complexity of evaluating each position is $n\cdotp m$, the total complexity is $n \cdotp m\cdotp\log n$. As the complexity of evaluating the impact of merging dominates the total complexity, the Linear Probing Heuristic, which spends less time evaluating the position, is more efficient.

\emph{Optimality Analysis.} Assume that there are multiple appropriate positions for task $i+j$. Logarithmic Probing Heuristic returns the first position it finds and meets the criteria. Thus, it is not biased to any certain appropriate position for the merged task. Alternatively, Linear Probing Heuristic always finds the latest appropriate position in the batch queue for task $i+j$. This ensures that task $i+j$ has the least impact on other tasks' completion times. Being the last possible position, however, increases the likelihood of $i+j$ to miss its deadline. In addition, this makes it unlikely for other tasks to be scheduled in front of $i+j$, hence, limiting the chance of future merging operations. 




\section{Adapting Task Merging based on the oversubscription level}
\label{sec:adapt}
\subsection{Overview}

In Section \ref{sec:appropriateness}, we discussed the merge appropriateness of each task by considering a worst-case analysis to assure no task is affected by the merging. However, when the system is oversubscribed, we can compromise the worst-case analysis and make the system more permissive to task merging in order to mitigate the oversubscription. In fact, sacrificing a few tasks in favor of more merging can lighten the system's oversubscription and ultimately cause fewer tasks to miss their deadlines. For that purpose, in this section, we develop the Workload Assessor component (see Figure~\ref{fig:adc}) that is in charge of assessing the oversubscription level of the system and accordingly adjusting the aggression level of applying the task merging. 


\subsection{Quantifying Oversubscription of a Computing System}

The level of oversubscription in the system can be quantified based on various factors, such as the rate of missing deadlines and the task arrival rate. The quantification can be achieved in a reactive manner (\ie from known metadata) or in a proactive manner (\ie based on the factors that suggest the system is about to get oversubscribed in the near future). In this part, we provide a method for Workload Assessor that uses decisive indicators of oversubscription to quantify the oversubscription level of a serverless computing system.


The first intuitive idea to quantify oversubscription is based on the (measured or expected) ratio of the task arrival rate to the processing rate \cite{mahato2018reliability}. In this case, a system is oversubscribed if it cannot process tasks as fast as it receives them. This idea has two main limitations: (A) It requires the knowledge of processing rate, which is difficult to accurately measure; (B) It is prone to report false negative in the oversubscription evaluation. In particular, it cannot discriminate between different circumstances that the ratio tends to one. Such a circumstance can occur when the tasks' arrival and processing rates are similar. However, the batch queue may be congested (\ie the system is oversubscribed) or may not be (\ie the system is not oversubscribed).

Another idea is to use the ratio of the number of tasks missing their deadlines to the total number of tasks executed \cite{mahato2018reliability}. This is based on the fact that an oversubscribed system cannot complete all its tasks on time. Thus, missing a high number of task deadlines suggests an oversubscribed situation. Although this idea has good potential, yet it falls short in quantifying the degree of oversubscription. That is, it cannot discriminate between a system that completes tasks a short time after their deadlines versus the one that completes tasks a long time after their deadlines.

Improving on the shortcomings of the aforementioned methods, we propose to quantify the oversubscription level of the system in a given time window based on the \emph{deadline miss severity ratio}. We define \emph{waitable time} of task $i$, denoted $W_i$, as the maximum time it can wait in the queue without missing its deadline. Let $A_i$ denote the arrival time of task $i$, then its waitable time is calculated as: $W_i = \delta_i-A_i- E_i$ . To quantify the oversubscription level, denoted $OSL$, in the first place, we discard the contribution of infeasible tasks (\ie those with $W_i < 0$) and those that can complete on time (\ie the ones with $C_i^m\leq\delta_i$). Next, the tasks that complete after their deadlines contribute to the oversubscription level based on the severity of their deadline miss. For a given task $i$, this is calculated based on the proximity between its completion time and its deadline (\ie $C_i^m-\delta_i$) and with respect to its waitable time (\ie $W_i$). Equation~\ref{eq:weightedmiss} formally shows how $OSL$ is calculated. Recall that $C_i^m$ is estimated based on Equation~\ref{eq:compl} to quantify the oversubscription in the current time window and $N_a$ represents the total number of tasks across all the machine queues. To adapt Equation \ref{eq:weightedmiss} for quantifying the oversubscription of a past time window, we need to replace the estimated completion time with the observed completion time of the tasks. 

 \begin{equation}\label{eq:weightedmiss}
 \vspace{-5pt}
  OSL = \frac{1}{N_a}\sum_{i=1}^{N_a} 
  \begin{cases}
  0,                     & W_i \le 0 \\
  0,                     & C_i^m \leq \delta_i \\
  \frac{C_i^m - \delta_i}{W_i} ,& C_i^m > \delta_i 
  \end{cases}
 \end{equation}

 
\subsection{Adaptive Task Merging Aggressiveness}
The method explained in Section~\ref{sec:appropriateness} estimates the side-effect of merging on other tasks in a conservative manner to assure that the merging does not cause their deadlines violated. 
In the face of oversubscription, estimation of the side-effect can be relaxed from the worst-case analysis to allow more aggressive task merging, hence, mitigating the oversubscription and increasing the overall QoS. 

To make the aggressiveness of task merging adaptive, based on the measured oversubscription level of the system, we modify the acceptable probability that a merge operation does not cause deadline violation on other tasks of the system. More specifically, for higher values of the oversubscription level, the acceptable probability that other tasks meet their deadlines should be diminished and vice versa. 
\reviewhl{
Therefore, we set the value of Standard Deviation coefficient ($\alpha$) to inversely scale against the $OSL$ value. We formulate $\alpha$ as $\alpha = \beta- 2 \cdotp \beta \cdotp OSL$ where $\beta$ is the maximum value of Standard Deviation coefficient.  The value of $\beta = 2$ allows $\alpha$ to scale in the range of [-2 , 2], which translates to real execution time being less than the estimated execution time with [2.3\% , 97.7\%] probability. That is, to consider task merging, without oversubscription ($OSL \rightarrow 0$), the system requires a high certainty (97.7\%) that a task completes on or before its estimated time. Conversely, under a high oversubscription ($OSL \rightarrow 1$),  merging can be enacted with a low on-time task completion probability (2.3\%).
}
\section{Performance Evaluation}

 \label{sec:evltn}
    

 \subsection{Experimental Setup}
We developed a prototype of the SMSE platform with the task merging mechanism in place. We made SMSE publicly available\footnote{https://github.com/hpcclab/adaptivemerging} for the research community and reproducibility purposes. In this study, to comprehensively examine various workloads with different configurations, we used SMSE in the emulation mode \reviewhl{(except for the first experiment)}. The task merging mechanism, proposed in this paper, is implemented as the Admission Control component of SMSE. \reviewhl{
For the sake of reproducibility, we modeled a serverless computing system to have the specifications equivalent to eight Small VMs in the Chameleon Cloud~\cite{keahey2019chameleon}. That is, we modelled the execution time of each task on each serverless machine based on the benchmarked execution time of that task on the Small VMs of the Chameleon Cloud.} 

The multimedia repository we used for evaluation includes multiple replicas of a set of benchmark videos. Videos in the benchmarking set are diverse both in terms of the content types and length. The length of the videos in the benchmark varies in the range of [10, 220] seconds, splitting into 5-110 video segments. The benchmark videos are publicly available for reproducibility purposes\footnote{\url{https://github.com/hpcclab/videostreamingBenchmark}}. More details about the characteristics of the benchmark videos can be found in our other study \cite{shangrui_paper}. For each segment of the benchmark videos, we obtained their execution times by executing each micro-service 30 times. The benchmarked micro-services are: reducing resolution, adjusting bit rate, adjusting frame rate, and changing codec. In each case, two conversion parameters are examined. For example, the frame rate is changed from 60 fps down to either 30 fps or 24 fps. Note that a codec changing micro-service can take up to 8x longer to execute than other more trivial processing operations \cite{shangrui_paper}.

To evaluate the system under various workload intensities, we generate [1,000, 2,500] video segment processing tasks within a fixed time interval. All transcoding micro-services are available in the processing units (\ie warm starting micro-services). Transcoding tasks arrive in the system in a group of 5 consecutive segments at a time. To accurately emulate common workload behavior observed in the real video steaming systems, each workload repeatedly toggle their arrival rate between the base period and high load period, where the arrival rate is increased by two folds. Each base period is approximately three times longer than the high load period. Each simulation case spans up to 15 high and base period cycles. 
In each simulation case, if all tasks arrive simultaneously, there is approximately 30\% chance for some tasks to find a mergeable pair. However, as the tasks are dynamically arriving in the system throughout the simulation time, the chance of task merging reduces to be less than 20\%. \reviewhl{These test cases are used throughout all the experiments except one in Section~\ref{subsec:saving0}.}

We collect the deadline miss-rate (DMR) and makespan (\ie execution time to finish all tasks) of completing all tasks.  \reviewhl{Although we employ ‘weighted estimated time after the deadline’ to fine-tune the response of adaptive merging based on the oversubscription level, we show DMR as the evaluation metric, which is a common metric to express the user satisfaction. DMR is an intuitive metric that makes the results comparable to other studies.
}
For the sake of better visualization of the miss rate reduction, the DMR of each configuration with merging policy is normalized against a nearly identical configuration without the task merging in place. We conducted each experiment 30 times, each time with different randomized task arrival time and order. Mean and 95\% confidence interval of the results is reported. In every experiment, all tasks must be completed, even if they miss their deadlines.

For each experiment, we examine the system in four scenarios: (A) Without task merging; (B) Conservative task merging policy (\ie by considering merge appropriateness to strictly not cause additional deadline miss); (C) Aggressive task merging policy (\ie without considering merge appropriateness); and (D) Adaptive task merging policy (\ie an adaptive system that works either similar to considerate or aggressive depending on the situation). However, for the sake of better presentation, only some parts of the results are shown in each experiment.



 \begin{figure}[b]
	\vspace{-10pt}
	\centering
	\includegraphics[width=0.4\textwidth]{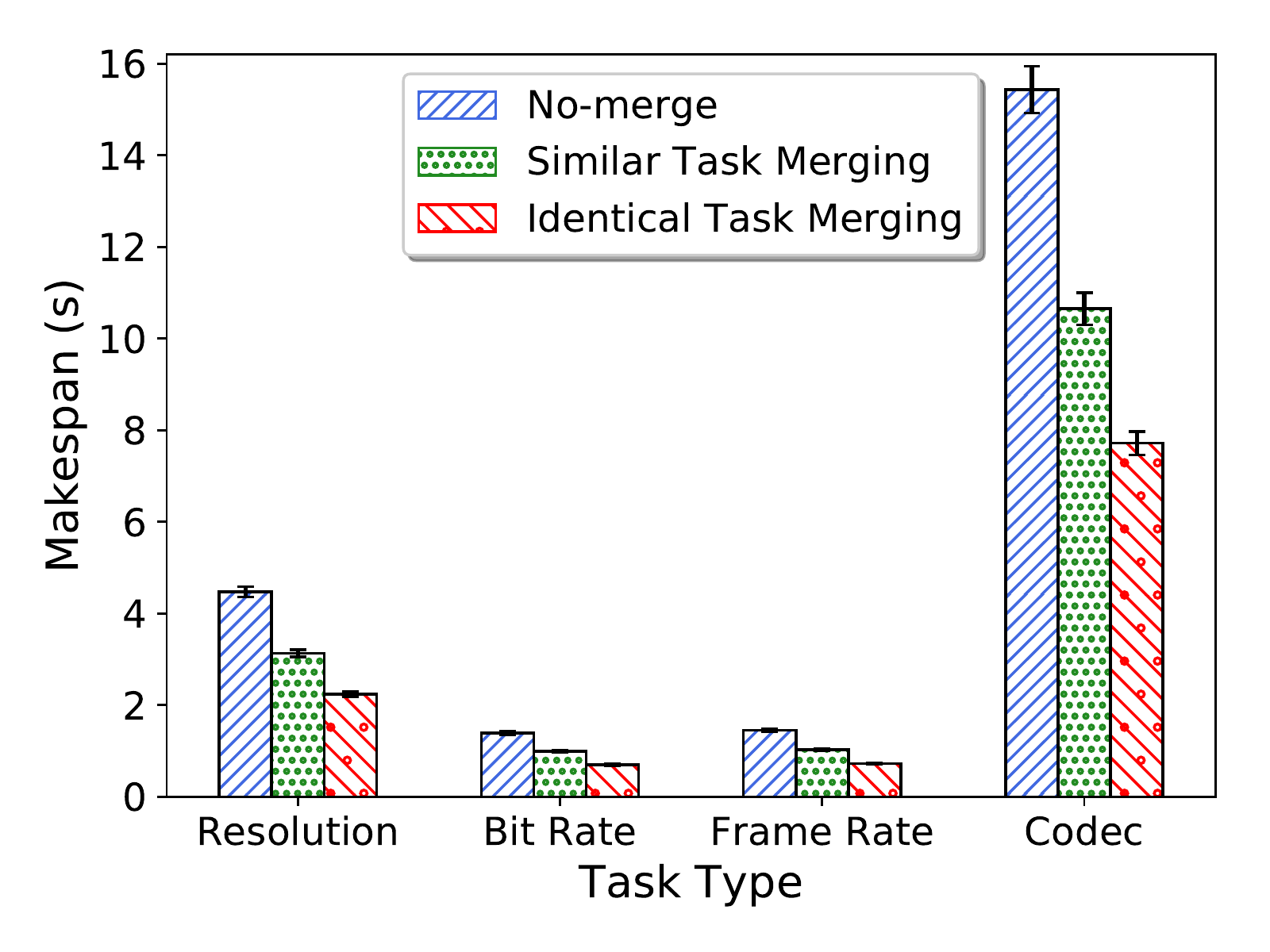} 
	\caption{\small{\reviewhl{Comparing the time to complete two tasks performing various video processing task types in three configurations: not merging, merging as similar tasks and merging as identical tasks.}}}
	\label{fig:saving1}%
	
\end{figure}

 \subsection{Evaluating the Resource Saving of Task Merging}
 \label{subsec:saving0}
\reviewhl{
In the first experiment, our goal is to examine the impact of task merging on the makespan time. For that purpose, we configured SMSE in the real-setting mode and executed two video tasks under the merged and unmerged scenarios. Specifically, we examined three cases: executing the tasks separately; merge them under data-only task merging; and merge them under the task level similarity.
}

\reviewhl{
As we can see in Figure~\ref{fig:saving1}, executing two tasks individually (\ie, without merging) takes approximately twice as much the time it takes to complete the tasks in the task level merging. Merging the tasks under data-only merging also saves around 30-45\% of the makespan time. This figure also shows the heterogeneity of task size in the system. Codec changing operation requires approximately eight times longer than the frame rate changing operation, for example.
}

 \subsection{Impact of the Task Merging on the Makespan Time}

In the next experiment, our goal is to see the impact of the task merging on makespan of the whole system. This metric implies the time cloud resources are deployed, and subsequently, the cost incurred to execute all the tasks. We examine the system under various subscription levels (from 1,000 to 2,500 requests) arriving within the time interval. In this experiment, we examine systems with three queuing policies, namely FCFS, EDF, and MU (Max Urgency). Also, the Position Finder component is disabled for this experiment. That means all the merged tasks are placed in the position of the existing task in the batch queue.
In this system, because tasks are not dropped and computing resources are homogeneous, the scheduling policies do not make a significant change in the makespan. Therefore, only the results of the FCFS queuing policy are presented. 

\begin{figure}[h]
	\centering
	\includegraphics[width=0.4\textwidth]{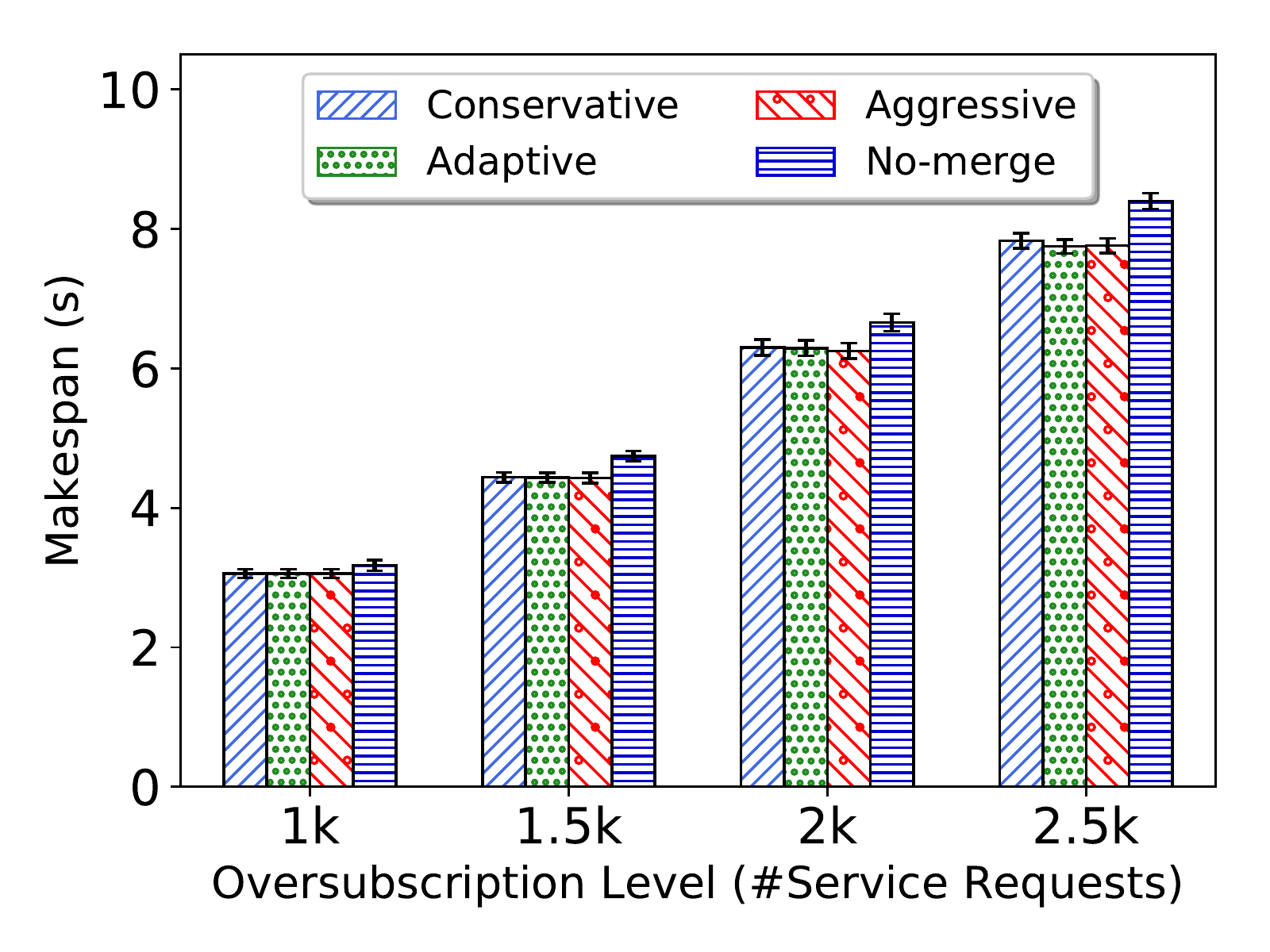} 
	\caption{\small{Comparing the total time to complete all tasks (\ie makespan) under a varying number of arriving processing tasks (horizontal axis) in four scenarios: without task merging, with Adaptive, Conservative, and Aggressive merging in place.}}
	\label{fig:exetime}%
\end{figure}

   \begin{figure*}[b] 
   \vspace{-10pt}
     \centering 
    
	\subfloat[ FCFS queue]{{\includegraphics[width=0.32\textwidth]{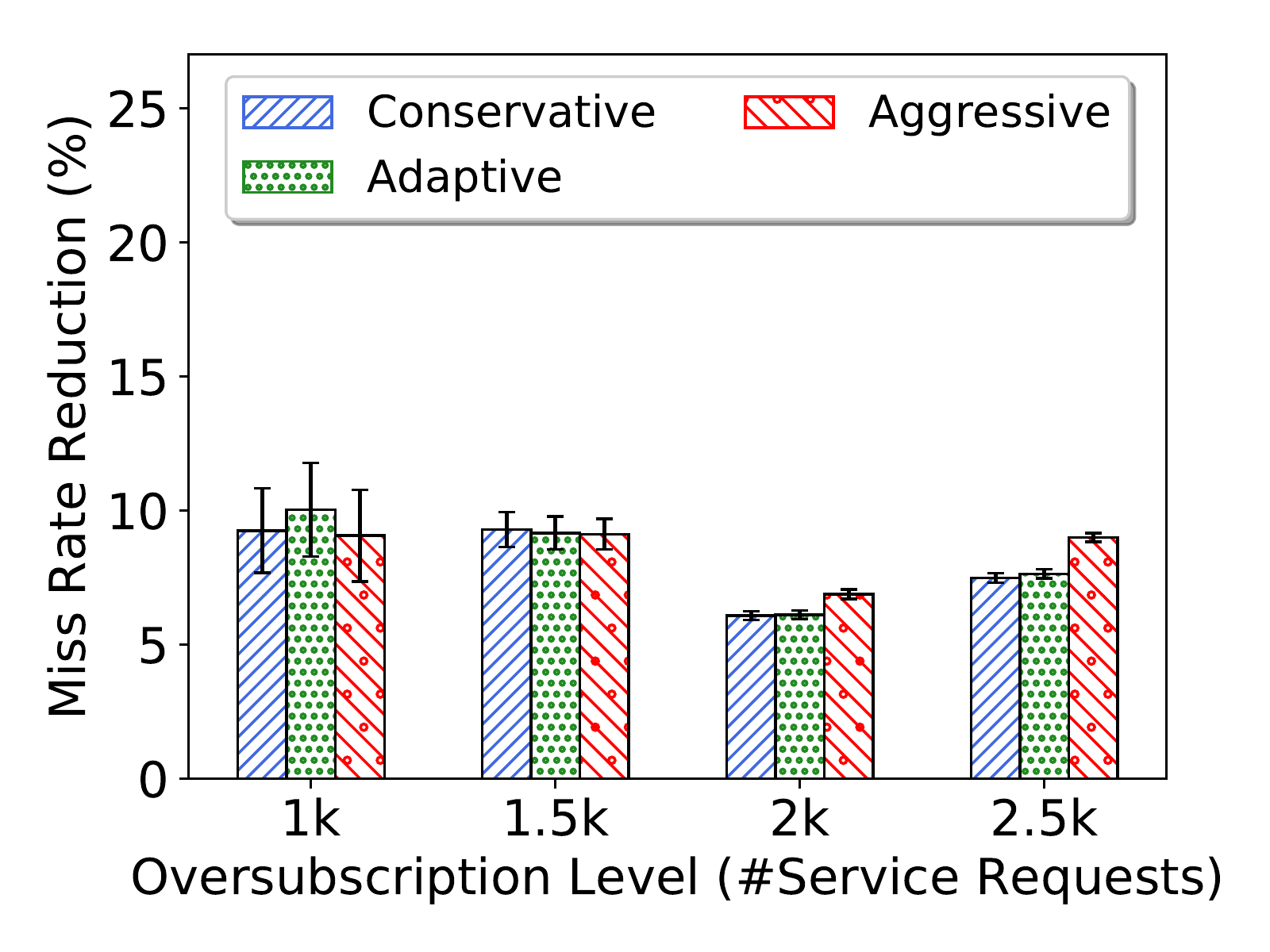} }} 
	\subfloat[ EDF  queue]{{\includegraphics[width=0.32\textwidth]{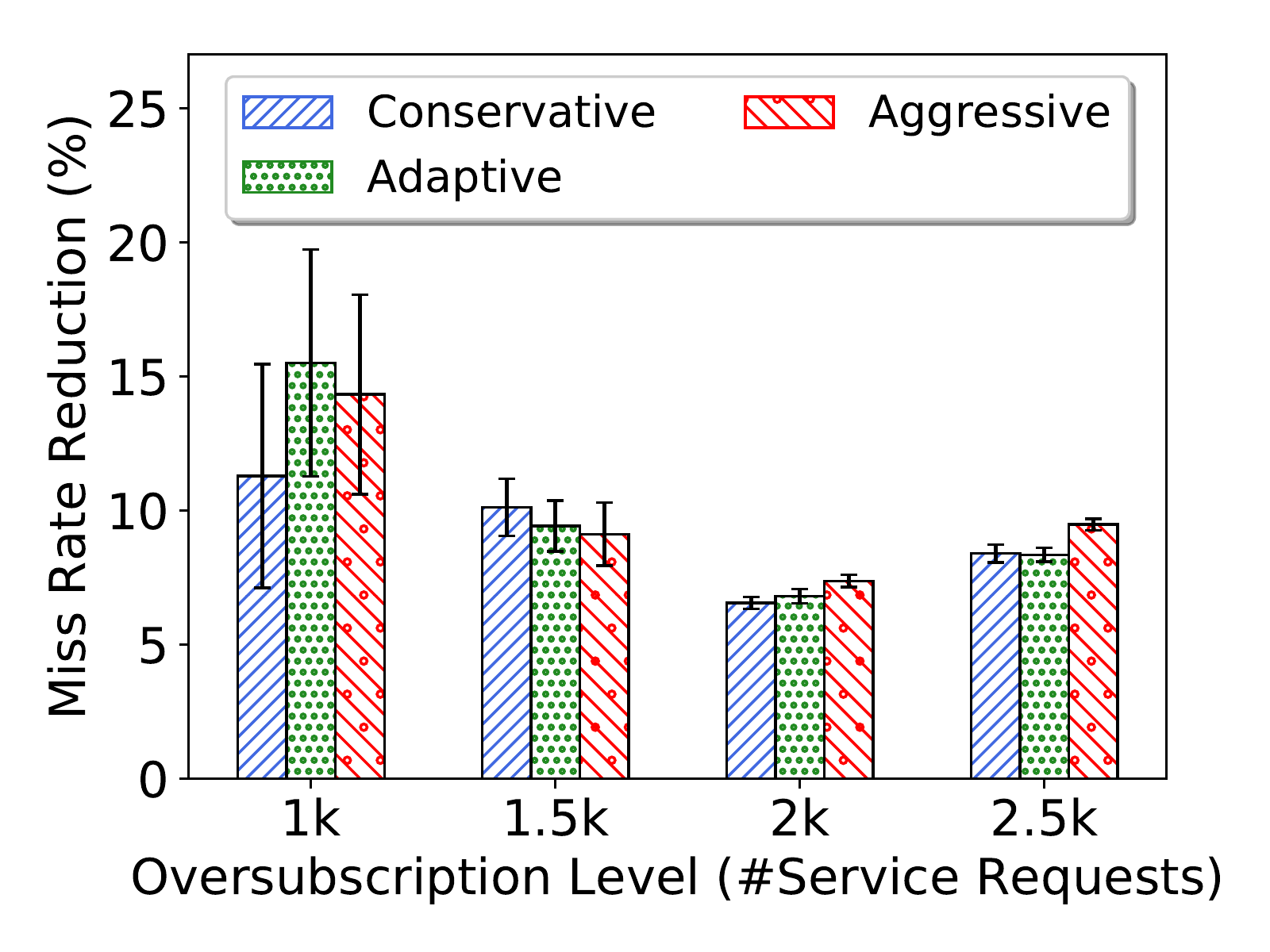} }} 
\subfloat[ MU queue]{{\includegraphics[width=0.32\textwidth]{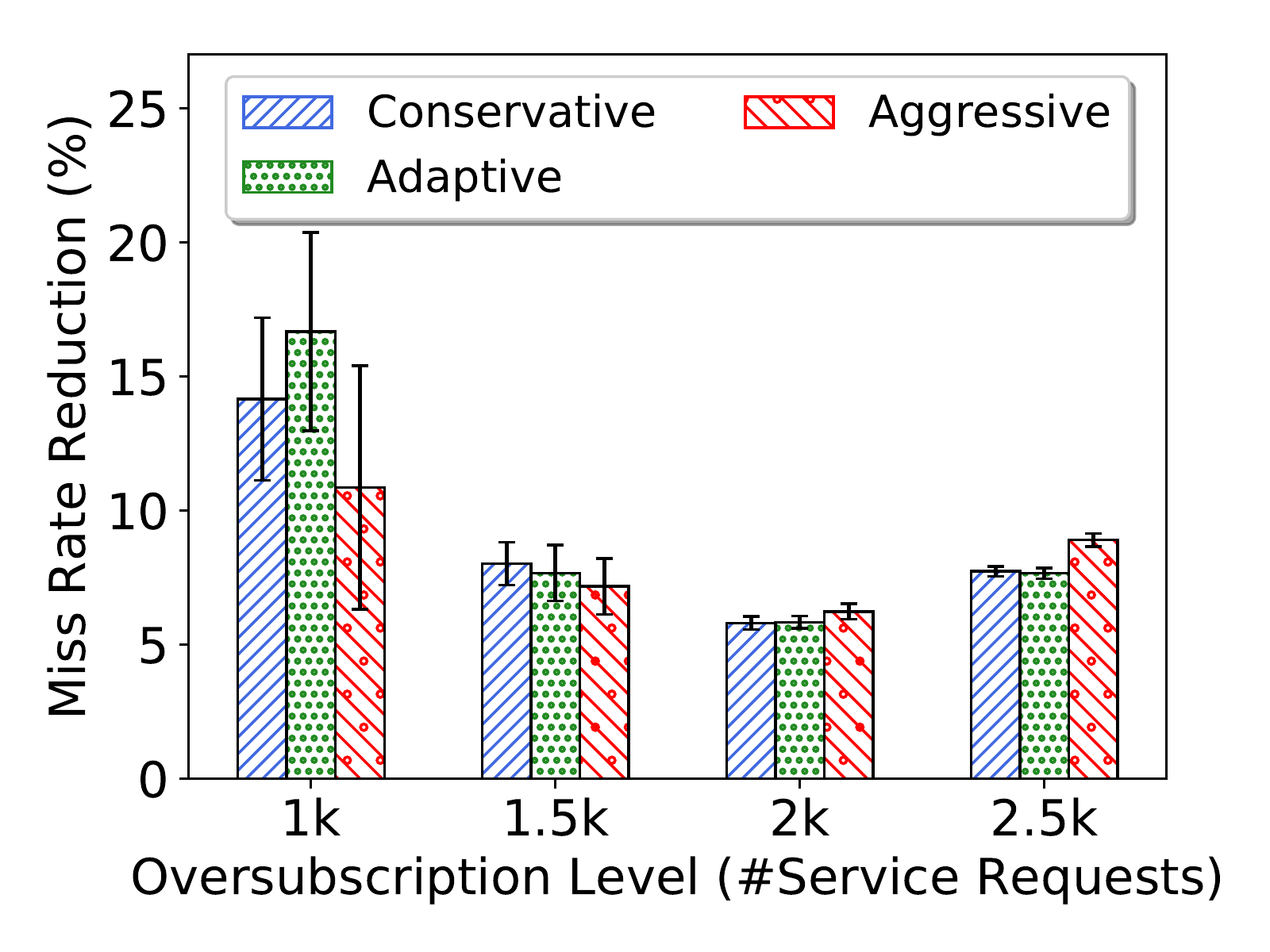} }}%
     \caption{\small{Comparing the deadline miss rate reduction under a varying number of tasks (horizontal axes) using Conservative, Aggressive, and Adaptive merging policies. Subfigures (a), (b), and (c) show the reduction under FCFS, EDF, and Max Urgency (MU) queuing policies.}}   
     \label{fig:deadline}%
 \end{figure*}

As we can see in Figure~\ref{fig:exetime}, our proposed merging mechanism saves the makespan between 4\% to 9.1\%. Saving in the makespan time is more pronounced when the system is highly oversubscribed. This is because there is more backlog of tasks in the scheduling queue at any moment. Hence, there is a higher chance of a new arriving task to find its mergeable pair. 

Comparing different task merging policies in the figure reveals that their difference in the total makespan is mostly marginal. The conservative merging policy is more reluctant to perform task merging, which can result in tasks missing their deadlines. However, that has an unanticipated positive effect on the total makespan by piling up more tasks in the early stage of its execution, which subsequently increases the chance of a new task to find a suitable mergeable pair at a later time.
Nonetheless, at a higher level of oversubscription, such effect is diminished, as there is a sufficient number of merge candidates in the batch queue regardless of the merging policy being employed. Thus, in a highly oversubscribed system, the makespan saving of the Conservative merging slightly lags behind other more Aggressive merging policies.

 \subsection{Impact of the Task Merging on QoS}
 In this experiment, our goal is to evaluate the viewers' QoS. For that purpose, we measure the deadline miss rate reduction resulted from merging tasks and compare them with the comparable systems that have no task merging under various oversubscribed levels. \reviewhl{Note that, in an oversubscribed system, the evaluated scheduling policies yield different DMR value. Nevertheless, the merging mechanism complements the employed scheduling policy and improve it to offer a better QoS (lower DMR value).} 
 
 As shown in Figure~\ref{fig:deadline}, we observe that task merging significantly reduces the deadline miss rate for all the scheduling policies. We observe that the improvement in the deadline miss rate of FCFS is less than the EDF and MU scheduling policies. \reviewhl{However, the reduction is more consistent across oversubscription levels when comparing to} the other policies. This is because FCFS, by nature, does not schedule tasks based on their deadlines \reviewhl{or urgencies. There is no critical point which the scheduler tip over from assigning mostly tasks with a tight but manageable deadline to mostly tasks with a too-short deadline.} 

The comparison across different merging policies reveals that, for low oversubscription levels, Conservative and Adaptive merging result in a higher deadline miss rate reduction than Aggressive merging. The reason is that the Aggressive merging makes inappropriate merging decisions that lead to deadline violation. However, as the oversubscription level increases, aggressively merging tasks seems to be the best approach \reviewhl{as the seemingly too aggressive task merging that yields immediate penalty results in less oversubscribed system later}.

Comparing the results shown in Figure~\ref{fig:exetime} with those in Figure~\ref{fig:deadline} reveals that the difference in deadline miss rate can be larger than the difference in makespan (\ie up to 18\% miss rate reduction compare to up to 9\% makespan reduction). This is because a small reduction in completion time can cause the merged tasks to meeting their deadlines instead of missing them. We can conclude that the impact of the task aggregation mechanism on viewers' QoS becomes more remarkable when it is combined with efficient scheduling policies.



 \subsection{Evaluating the Impact of the Position Finder}
 In this part, we examine the effect of the merge position finder module from Section~\ref{sec:posrespect} on the deadline miss rate reduction. We assume the system to schedule tasks in the FSCS manner while each of the merged tasks has a chance to be placed outside of their original order in the queue (using Linear Probing heuristic). We apply different merging policies without and with the position finder module (represented as \texttt{+Pfind} in 
 the figure) 
 in place.
 
   \begin{figure}[h]
	\centering
	\vspace{-10pt}
	\includegraphics[width=0.4\textwidth]{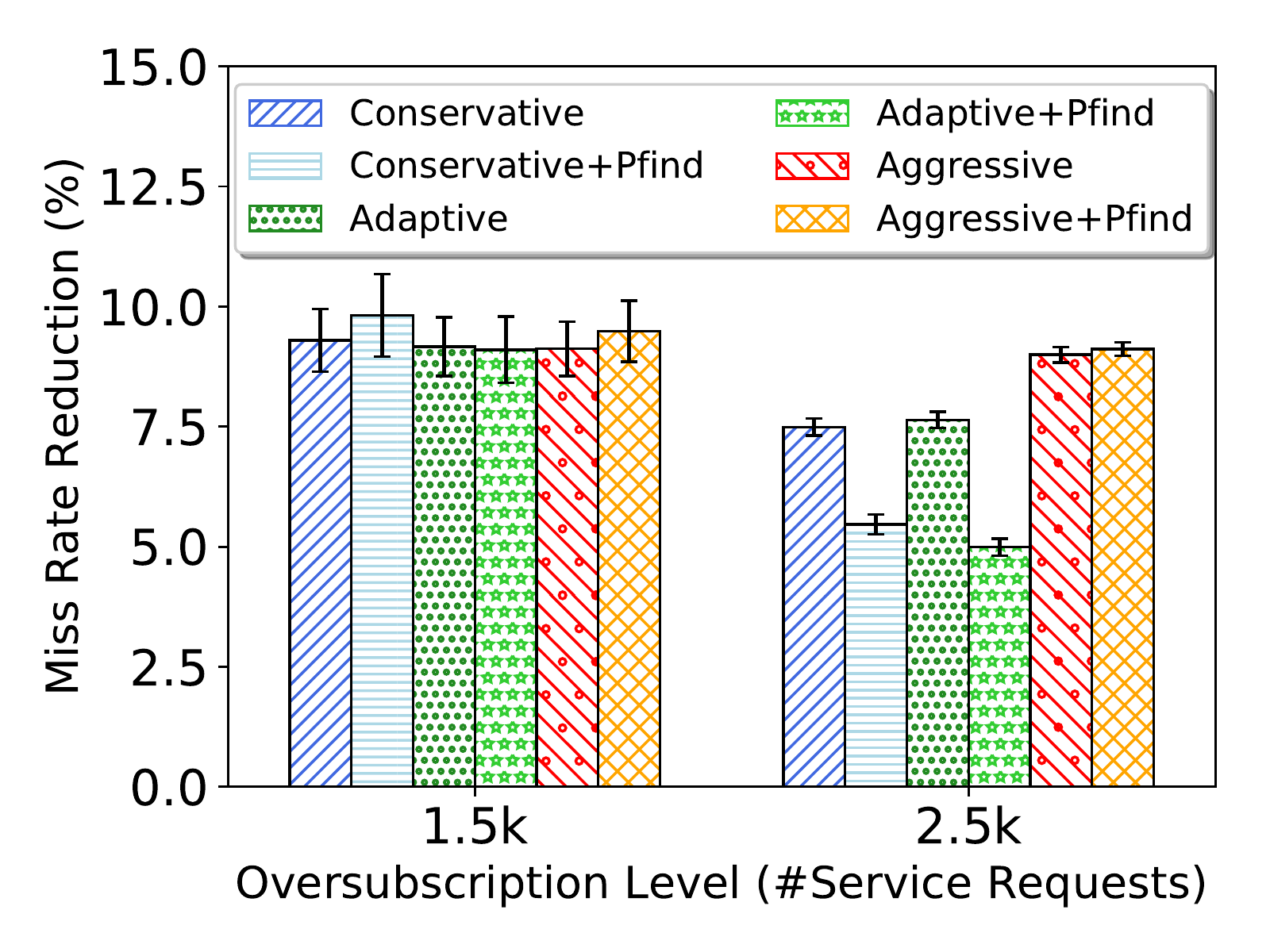} 
	\caption{\small{Comparing the effect of position finder module in term of deadline miss rate reduction under a varying number of arriving tasks (horizontal axis) with three merge aggressiveness levels that are applied without and with the position finder, shown as \texttt{Pfind} in the chart. }}%
	\label{fig:Pfind}%
	\vspace{-5pt}
\end{figure}

 Figure~\ref{fig:Pfind} shows an interesting result where the position finding module not only improves the deadline miss rate reduction in most cases but also degrades the performance of Conservative and Adaptive merging policies under highly oversubscribed conditions. This is because the position finder module places each merged task in a position that introduces the least amount of impact on other tasks. However, such a position that puts the least amount of impact on other tasks is also a position that is the closest to miss its own deadline. This, at the edge position, limits future mergeability should the other tasks want to merge in front of it. This is not the case for a system with a low task arrival rate because it is likely that the merged task completes its execution before another task merging in front of it. Also, Aggressive merging does not concern with the merge appropriateness. Thus, future mergeability is not reduced by the merged task placement. Accordingly, we recommend against using the Position Finder module with the Conservative and Adaptive merging policies in the face of high oversubscription levels.

 \subsection{Impact of the Execution Time Uncertainty on Task Merging}
As we noticed in Section~\ref{sec:impact}, merging decisions are made based on their impacts on the completion time distribution of other tasks. However, the magnitude of uncertainty in the execution time distribution of the tasks can be a decisive factor in the accuracy of estimating the merging side-effects, and subsequently, the deadline miss rate resulted from them. Accordingly, in this experiment, our aim is to evaluate how the three task merging policies function in the face of different levels of uncertainty in the execution time. For that purpose, we increase the randomness of execution time when sampling from the mean execution times. The base level of uncertainty in execution time distribution, observed from the video transcoding services, is relatively low, as the standard deviation is approximately 4\% of the mean execution time. In this part, we examine the deadline miss reduction when the standard deviation of execution time distribution is increased by 5 and 10 times, expressed as \texttt{5SD} and \texttt{10SD} in the results, under different oversubscription levels in the system. 

 \begin{figure}[t]
 \vspace{-10pt}
	\centering
	\includegraphics[width=0.4\textwidth]{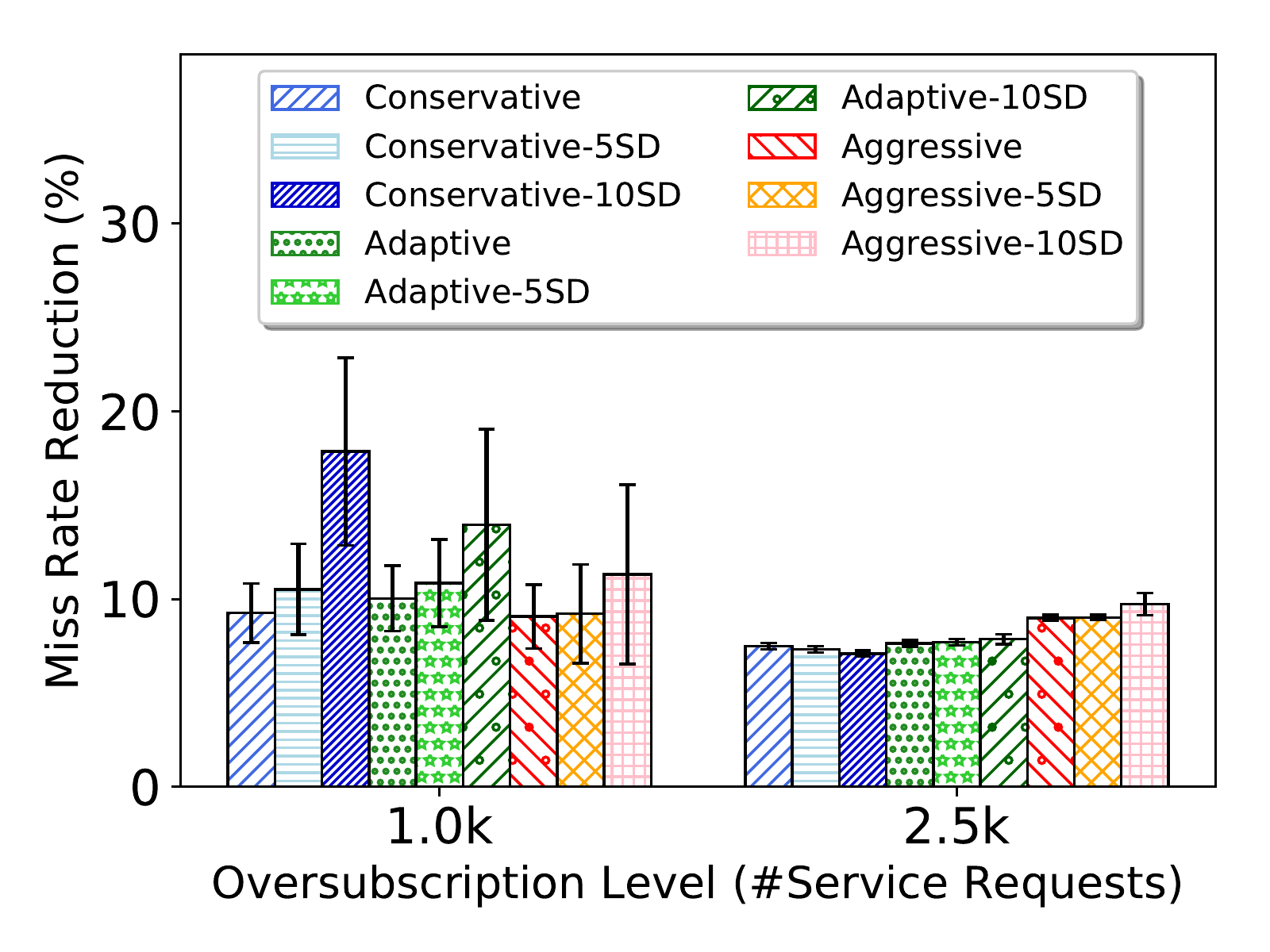} 
	\caption{\small{Comparing the deadline miss rate reduction for the different number of arriving tasks (horizontal axis) using the three merging policies applied on tasks with different uncertainty in their execution time distribution. \texttt{5SD} and \texttt{10SD} designate five times and ten times the randomness than the regular dataset. }}%
	\label{fig:varySD}%
	\vspace{-10pt}
\end{figure}
 
  \begin{figure}[h]
	\centering
	\includegraphics[width=0.4\textwidth]{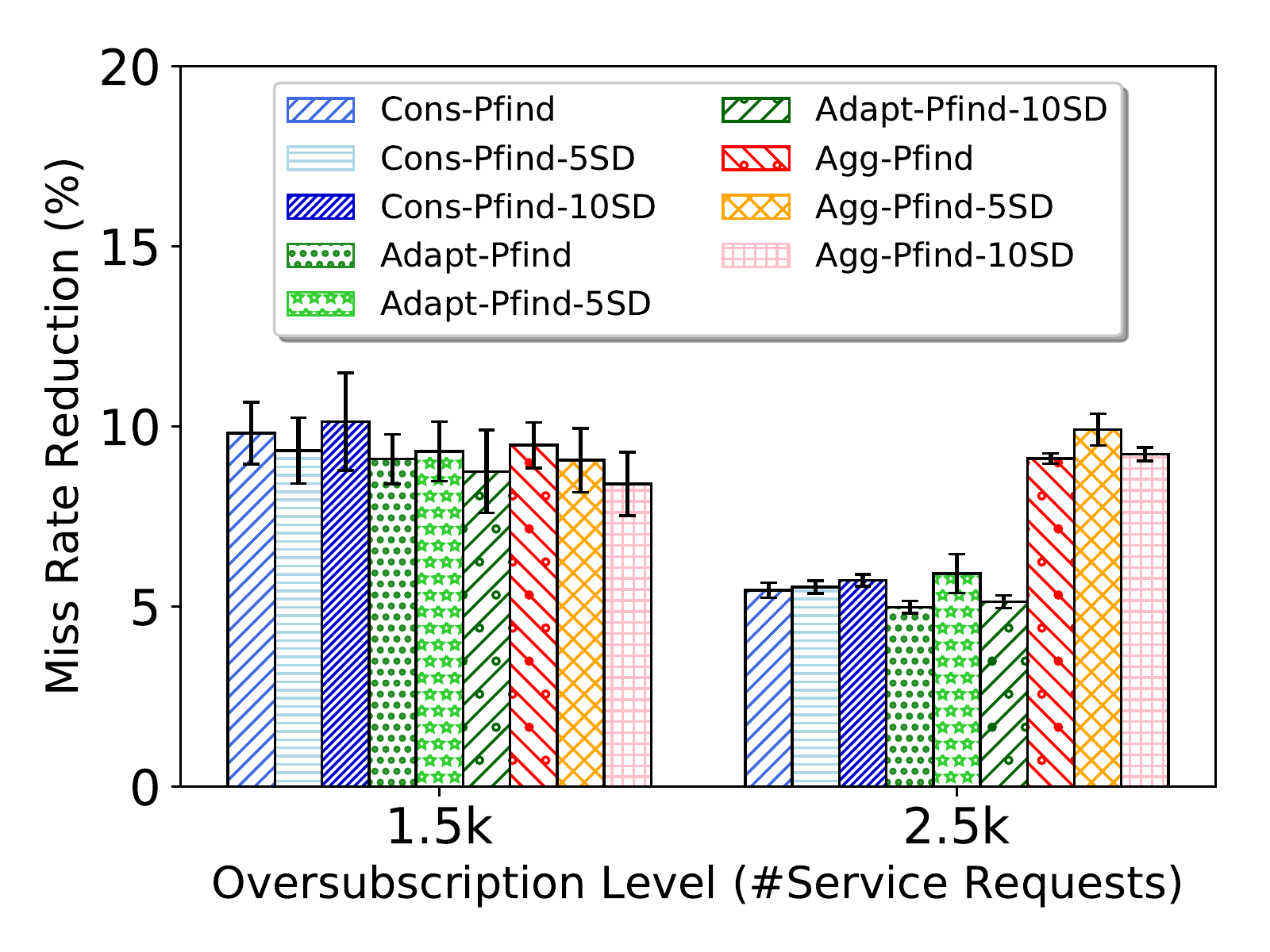} 
	\caption{\small{Comparing the deadline miss rate reduction under a varying number of arriving tasks (horizontal axis) using the three merging policies and three levels of execution time uncertainty. In every case, the position finder module (\texttt{Pfind} in the chart) is activated. \texttt{5SD} and \texttt{10SD} designate five and ten times more uncertainty in execution time distribution than the regular workload trace. }}%
	\label{fig:varySDPOS}%
\end{figure}

The result of this experiment, shown in Figure~\ref{fig:varySD}, includes some interesting observations. Specifically, we observe that as the level of uncertainty rises, there is more performance gain in performing merging. At the low oversubscription level, Conservative and Adaptive merging policies, both of which consider the standard deviation coefficient ($\alpha$) and the merge impact evaluation, gain more deadline miss reduction than the Aggressive policy. 
However, at a high oversubscription level (2.5k) with a high level of uncertainty, unlike other merging policies, Conservative merging often evaluates merging options as too risky to impacting other tasks. Therefore, the Conservative merging does not allow as many task merging as other policies, and thus performance gain is reduced as the uncertainty level rises. Adaptive merging does not exhibit such behavior and perform well in both situations.

Figure~\ref{fig:varySDPOS} shows the impact of increasing the uncertainty level on the performance of the position finder module. Comparing the result to those from Figure~\ref{fig:varySD}, we learn that when the position finder module is engaged, the increasing level of uncertainty only has a minimal impact on the deadline miss rate reduction. At the 2.5k oversubscription level, the Aggressive policy with the help of the position finder module still performs significantly better than other policies. 



\section{Related Works}\label{sec:relwk}
Software-based computational reuse has been extensively researched and deployed. However, not many systems can merge and reuse tasks before they are completed, and many of them tie very closely to one specific application. Below are some notable works in this area.

Chard~\etal a FaaS cloud platform named funcX~\cite{chard2020funcx}. FuncX containerizes functions to be executed at various endpoints most suitable to the function nature (\eg high CPU power machines or low storage latency machines close to the data source). Their solution can reuse the computing result that matches the recently executed task through the caching system. 

Elgamal~\emph{et al.}~\cite{elgamal2018costless} proposed a serverless computing platform with the capability to fuse (\ie merge) functions together to reduce the incurred cost of using the serverless computing cloud. Unlike our work, which merge unrelated tasks to reuse some part of the computation based on some similarity, their work fuse functions of the same workflow to reduce the data transition cost. Similar to our work, not all possible fusing options are appropriate. They evaluate the trade-off between fusing functions together and treat them separately through estimated incurring costs on a task by task basis. 

Popa~\emph{et al.}~\cite{DryadInc} presented modules to identify identical and similar tasks to cache partial results and reuse them on incremental computation specifically on Dryad \cite{Dryad} platform context. 
They proposed two solutions: One solution automatically caches computational results. Another solution merges tasks based on the programmer's defined merge function. Their first solution is a caching system, while their second solution is a computational reuse system specific to the Dryad platform, which does not have a deadline and QoE to consider.

Gunda~\emph{et al.}~\cite{Nectar} proposed an architecture that instead of caching finished computation product, it offers caching of intermediate computed data, which can derive into multiple versions of finished results with minimal further computation. They treated data and computation interchangeably in the sense that data can be replaced by re-computation to regain that data. Hence they could cache more intermediate computational data and less final results to save storage. However, they do not merge tasks and plan the scheduler to reuse intermediate data that will be available but currently is not ready.

Tiwari and Solihin~\cite{tiwari2014mapreuse} proposed a module to deduplicate identical data and merging tasks in the MapReduce platform using IMMR (In-Memory-MapReduce). Unlike much other deduplication in storage system based MapReduce, merging data in memory based MapReduce requires merging operations to perform at the data-structure level instead of the file system, and all operations must be done in a more timely manner. This is an eminent work showing the deduplication system on MapReduce style applications.  

Tang and Yang~\cite{SecureDedup} proposed a system to deduplicate storage and intermediate computational data among multiple mutually distrusted users in a cloud computing system in a secure way through API call and kernel modules. They use Radis for caching works, which they isolate cache by application signature in a way that only the same applications can share data together. The solution is especially helpful for libraries and services that are commonly used among multiple users. However, those libraries and services need to be modified to use their work's caching API. They focus on providing a securely shared key-stored caching system that allows coders to freely select when and what intermediate computation to cache through API, unlike our works, which focus more on detect and merging tasks with deadline consideration.

Boos \emph{et al.}~\cite{boos2016flashback} exploited pre-caching technique to mitigate real-time VR rendering performance issue on low-power mobile GPU. This work is focused on how to do computational reuse of similar scenes and rendering tasks in a Virtual Reality context and not applicable to other fields.

Samadi \emph{et al.}~\cite{samadi2014paraprox} proposed a system that identifies potentially computationally reusable tasks and uses them for software-only $approximate$ computation (\ie reusing the computations that would give close enough results rather than re-processing to get accurate results) to improve performance, especially in Video processing context where accurate computational results are not required. Unlike our work, they heavily prefer performance improvement over the user's QoS, who is now getting inaccurate-results.

Paulo and Pereira \emph{et al.}~\cite{paulo2014distributed}
developed a system to perform deduplication of high throughput data using $Bloom filters$. $Bloom filters$, while fast, has chances of giving false positive hash checking. Therefore they achieve lower overhead data duplicate detection than the hash-table approach we use, at the price of compromised accuracy.
Marahatta \emph{et al.}~\cite{marahatta2019classification} propose a  dynamic task scheduling that emphasizes energy efficiency of the cloud data center. Part of their solution utilizes task merging to reduce mean response time and total energy consumption. Unlike our work, where we focus on oversubscribed conditions, their work aims to lower the energy consumption of underutilized data center.

\section{Conclusion and Future works}\label{sec:conclsn}
In this paper, we investigated the problem of oversubscription in the serverless computing platforms. Our goal is to alleviate the oversubscription via merging arriving service requests (task) with other (exact or similar) tasks in the system. In that regard, we dealt with two challenges: \emph{First}, how to identify identical and similar tasks in an efficient manner? \emph{Second}, how to perform (or not perform) merging to achieve the best QoS in the system? 
To address the first challenge, we identified three main levels of similarity that tasks can be merged. Then, we developed a method to detect different levels of task similarity within a constant time complexity. To address the second challenge, we developed a method that determines, based on system oversubscription condition, how to perform the merge operation so that the deadlines of other tasks in the system are likely least affected. Experimental results demonstrate that task merging can reduce the overall execution time of tasks by more than 9\%. Hence, cloud resources can be deployed for a shorter time. Interestingly, this benefit comes with improving the QoS of the users by up to 18\%. We concluded that when the level of oversubscription in the system is high, merging tasks aggressively (\ie without being considerate of the impact on other tasks) helps in improving the QoS. Conversely, with lower levels of oversubscription, merging should be carried out with consideration of the impact on other tasks, not to cause an unnecessary impact on the QoS.

Although we implemented this system in the context of a video streaming platform, the concept can be applied to the computing platform of other domains as long as we can define similarity levels in those domains.
In the future, we plan to extend this work by considering the impact of using heterogeneous computing resources in the system. 
\reviewhl{Another interesting future research direction can be exploring the impact of marginally compromising tasks' specification accuracy (substitute some task parameters with similar value) to enable more computational sharing with other existing tasks. Such an approach requires semantic similarity (as opposed to deterministic similarity) detection that operates based on machine learning. However, as machine learning methods impose considerably more overhead than the hash-table approach, the system must take into the account the cost benefit of utilizing the machine learning before employing them.} 


\section*{Acknowledgments}
We would like to thank the anonymous reviewers of the paper. This is a substantially extended version of a paper presented at the 16th International Conference on Service-Oriented Computing (ICSOC '18) \cite{icsoc18}. This research is supported by the National Science Foundation under award\# CNS-2007209 and CNS-2047144.

\bibliographystyle{IEEEtran}

\bibliography{references}

\begin{IEEEbiography}[{\includegraphics[width=1in,height=1.15in,clip,keepaspectratio]{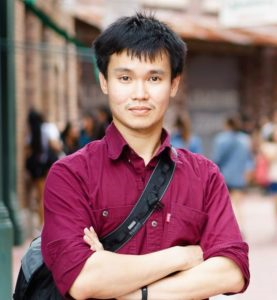}}]{Chavit Denninnart}
is a Ph.D. student at High Performance Cloud Computing (HPCC) laboratory at University of Louisiana at Lafayette. His research interests include cloud based resource allocation, serverless computing, computational reuse, heterogeneous computing, and low-latency cloud-based video streaming. 
\end{IEEEbiography}

\begin{IEEEbiography}[{\includegraphics[width=1in,height=1.25in,clip,keepaspectratio]{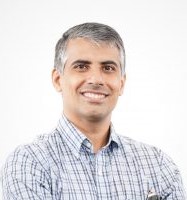}}]{Mohsen Amini Salehi} 
received his Ph.D. in Computing and Information Systems from Melbourne University, in 2012. He is an NSF CAREER Awardee Assistant Professor and the director of HPCC lab, at the School of Computing and Informatics, University of Louisiana at Lafayette. His research focus is on edge-to-cloud continuum, heterogeneity, virtualization, resource allocation, energy-efficiency, and security.
\end{IEEEbiography}
\end{document}